\documentclass[prd,preprint,superscriptaddress,showpacs,byrevtex]{revtex4}
\usepackage{bm}
\def\fsl#1{\setbox0=\hbox{$#1$}           
   \dimen0=\wd0                                 
   \setbox1=\hbox{/} \dimen1=\wd1               
   \ifdim\dimen0>\dimen1                        
      \rlap{\hbox to \dimen0{\hfil/\hfil}}      
      #1                                        
   \else                                        
      \rlap{\hbox to \dimen1{\hfil$#1$\hfil}}   
      /                                         
   \fi}                                         %
\newcommand{\be}{\begin{equation}}
\newcommand{\ee}{\end{equation}}
\newcommand{\bea}{\begin{eqnarray}}
\newcommand{\eea}{\end{eqnarray}}
\newcommand{\beq}{\begin{equation}}
\newcommand{\eeq}{\end{equation}}
\newcommand{\beqs}{\begin{eqnarray}}
\newcommand{\eeqs}{\end{eqnarray}}

\begin{document}
\title{ Proof of Color Octet NRQCD Factorization of P-Wave Heavy Quarkonium Production at All Orders in Coupling Constant }
\author{Gouranga C Nayak }\thanks{E-Mail: nayakg138@gmail.com}
\date{\today}
\begin{abstract}
Recently we have proved color octet NRQCD factorization of S-wave heavy quarkonium production at all orders in coupling constant at high energy colliders in \cite{nayaknr}. In this paper we extend this to prove color octet NRQCD factorization of P-wave heavy quarkonium production at all orders in coupling constant at high energy colliders. We find that while the color octet NRQCD S-wave non-perturbative matrix element contains two gauge-links in the adjoint representation of SU(3), the color octet NRQCD P-wave  non-perturbative matrix element contains four gauge-links in the fundamental representation of SU(3).
\end{abstract}
\maketitle
\pagestyle{plain}
\pagenumbering{arabic}
\section{Introduction}

There has been lot of progress in the NRQCD heavy quarkonium production \cite{nayaknr,nrqcd,allnrqcd} to explain the experimental data at the Tevatron \cite{tev} and at the LHC \cite{lhc}. In the original formulation of NRQCD \cite{nrqcd} the proof of factorization of heavy quarkonium production at the high energy colliders was missing. The factorization theorem plays an important role at high energy colliders to study physical observables \cite{cs,nayakst,nayaknr,nkstjq,nkjhp}.

In the original formulation of NRQCD heavy quarkonium production the non-perturbative matrix element was given by \cite{nrqcd}
\bea
{\cal O}_H = <\Omega|\eta^\dagger(0) L_n \zeta(0) a^\dagger_H a_H \zeta^\dagger(0) L'_n \eta(0) |\Omega>
\label{naq}
\eea
which is not gauge invariant and is not consistent with the factorization of infrared (IR) divergence \cite{nayakst} where $\eta ~(\zeta)$ is the two component spinor field that creates (annihilates) a heavy quark, $a^\dagger_H$ is the creation operator of the heavy quarkonium and $L_n$ contains factors such as color matrix $T^a$ and derivative operators etc.. In eq. (\ref{naq}) the $|\Omega>$ is the full interacting (non-perturbative) vacuum which is different from the perturbative vacuum $|0>$.

The proof of NRQCD factorization of heavy quarkonium production at next-to-next leading order (NNLO) in coupling constant was given in \cite{nayakst} by using the diagrammatic approach where it was found that the gauge links were needed in the non-perturbative matrix element to make it gauge invariant and be consistent with the factorization of infrared (IR) divergence. The gauge invariant definition of the color octet NRQCD S-wave non-perturbative matrix element of heavy quarkonium production consistent with the factorization of infrared divergence at NNLO in coupling constant is given by \cite{nayakst}
\bea
{\cal O}_H = <\Omega|\eta^\dagger(0) T^a \zeta(0) \Phi^\dagger_{adj}(0) a^\dagger_H a_H \Phi_{adj}(0)\zeta^\dagger(0) T^a \eta(0) |\Omega>
\label{anaq}
\eea
where $\Phi_{adj}(x)$ is the light-like gauge link in the adjoint representation of SU(3) given by
\bea
\Phi_{adj}(x)={\cal P} e^{-ig T^d_{adj} \int_0^\infty dz l\cdot A^d(x+lz)},~~~~~~~~~~~~~~l^2=0.
\label{bnaq}
\eea
Recently we have proved the color octet NRQCD factorization of S-wave heavy quarkonium production at all orders in coupling constant at high energy colliders in \cite{nayaknr} by using the path integral formulation of QCD. We have found that the gauge invariant definition of the color octet NRQCD non-perturbative matrix element of S-wave heavy quarkonium production consistent with the factorization of infrared divergence at all orders in coupling constant is given by eq. (\ref{anaq}) \cite{nayaknr}. This confirms that the gauge invariant definition of the color octet NRQCD S-wave non-perturbative matrix element in eq. (\ref{anaq}) which was predicted by the NNLO diagrammatic calculation of factorization of infrared divergence in \cite{nayakst} agrees with the corresponding prediction from the factorization of infrared divergence at all orders in coupling constant by using the path integral formulation \cite{nayaknr}.

In this paper we extend this to prove color octet NRQCD factorization of P-wave heavy quarkonium production at all orders in coupling constant at high energy colliders. We find that while the color octet NRQCD S-wave non-perturbative matrix element in eq. (\ref{anaq}) contains two gauge-links in the adjoint representation of SU(3), the color octet NRQCD P-wave  non-perturbative matrix element contains four gauge-links in the fundamental representation of SU(3).

We find that the gauge invariant definition of the color octet NRQCD non-perturbative matrix element of P-wave heavy quarkonium production consistent with the factorization of infrared divergence at all orders in coupling constant is given by
\bea
{\cal O}_H = <\Omega| \eta^\dagger(0) \Phi(0){\overline { \nabla}} T^a \Phi^\dagger(0) \zeta(0) a^\dagger_H \cdot a_H \zeta^\dagger(0) \Phi(0) T^a{\overline { \nabla}}  \Phi^\dagger(0) \eta(0) |\Omega>
\label{dnaq}
\eea
where $\Phi(x)$ is the light-like gauge link in the fundamental representation of SU(3) given by
\bea
\Phi(x)={\cal P} e^{-ig T^d \int_0^\infty dz l\cdot A^d(x+lz)},~~~~~~~~~~~~~~l^2=0
\label{enaq}
\eea
and ${\overline { \nabla}}$ is defined by
\bea
\eta {\overline { \nabla}} \zeta =\eta ({ {\vec \nabla}} \zeta) -( { {\vec \nabla}}\eta) \zeta.
\label{fnaq}
\eea

We find that the gauge invariant definition of the color octet NRQCD non-perturbative matrix element of P-wave heavy quarkonium production consistent with the factorization of infrared divergence at all orders in coupling constant in eq. (\ref{dnaq}) is independent of the light-like vector $l^\mu$ used to define the light-like gauge link $\Phi(x)$ in eq. (\ref{enaq}).

We will provide a proof of eq. (\ref{dnaq}) in this paper.

The paper is organized as follows. In section II we discuss the infrared divergence in quantum
field theory and the light-like eikonal line. In section III we prove color octet NRQCD factorization of
P-wave heavy quarkonium production at all orders in coupling constant. In section
IV we derive the gauge invariant definition of the color octet NRQCD non-perturbative matrix element of P-wave heavy quarkonium production consistent with the factorization of infrared divergence at all orders in coupling constant. Section V contains conclusions.

\section{ Infrared Divergence in Quantum Field Theory And The Light-Like Eikonal Line }

In this section we will discuss the infrared (IR) divergence in quantum field theory and the light-like eikonal line. For simplicity we will consider the QED situation first before considering the QCD situation. We will focus on the infrared divergence arising due to the photon interaction with the light-like eikonal line. We will show that the infrared divergences due to the photon interaction with the light-like eikonal line in the eikonal approximation can be described by the pure gauge field. This enormously simplifies the study of factorization of infrared divergences due to the presence of light-like eikonal line in the quantum field theory at all orders in coupling constant.

Infrared divergence in QED occurs in the real photon emission from a single electron (or from single positron) and in the virtual photon exchange between electron-electron (or electron-positron or positron-positron) pair.

\subsection{ Real Photon Emission From Single Electron and The Infrared Divergence }

For a real photon of momentum $k^\nu$ emitted from the electron of momentum $k_1^\nu$ the Feynman diagram contribution is given by
\bea
{\cal A}=\frac{1}{{\not k}_1-{\not k} -m_e} {\not \epsilon}(k)u(k_1)
\label{ek1}
\eea
which can be written as
\bea
{\cal A}={\cal A}_{eik}+{\cal A}_{non-eik}
\label{ek2}
\eea
where
\bea
{\cal A}_{eik}=-\frac{k_1 \cdot \epsilon(k)}{k_1 \cdot k}u(k_1),~~~~~~~~{\cal A}_{non-eik}=\frac{{\not k} {\not \epsilon}(k)}{2k_1 \cdot k}u(k_1).
\label{3ek}
\eea
Decomposing the photon field $\epsilon_\nu$ as sum of pure gauge field $\epsilon_\nu^{pure-gauge}$ plus the physical field $\epsilon_\nu^{physical}$ we write $\epsilon_\nu=\epsilon_\nu^{physical}+\epsilon_\nu^{pure-gauge}$ where \cite{gy}
\bea
\epsilon_\nu^{physical}=\epsilon_\nu -k_\nu \frac{k_1 \cdot \epsilon(k)}{k_1 \cdot k},~~~~~~~~~~~~\epsilon_\nu^{pure-gauge}=k_\nu \frac{k_1 \cdot \epsilon(k)}{k_1 \cdot k}.
\label{4ek}
\eea
Using eq. (\ref{4ek}) in (\ref{3ek}) we find in the infrared limit (in the limit $k \rightarrow 0$)
\bea
&&{\cal A}_{eik}^{physical}=0,~~~~~{\cal A}_{non-eik}^{pure-gauge}=0,~~~~~~~~{\cal A}_{non-eik}^{physical}={\rm finite},~~~~~~ {\cal A}_{eik}^{pure-gauge}\rightarrow \infty.\nonumber \\
\label{5ek}
\eea
From eq. (\ref{5ek}) we find that if the photon field is the pure gauge field then the infrared divergence in the quantum field theory can be studied by using the eikonal approximation without modifying the finite part of the cross section.

Since the light-like eikonal line produces pure gauge field in quantum field theory, see subsection \ref{eikpg}, we find that the factorization of the infrared divergence in quantum field theory due to the presence of the light-like eikonal line is enormously simplified by using the pure gauge field.

\subsection{ Virtual Photon Exchange Between Electron-Electron Pair and The Infrared Divergence }

In the previous subsection we have considered the situation where the photon interacts with the single electron. In this subsection we will consider the situation where the photon interacts with more than one electron. Note that when the photon interacts with the single electron then that photon is the real photon but when the photon interacts with more than one electron then that photon is the virtual photon.

Let the momenta of two electrons be $k_1^\nu$ and $k_2^\nu$ and the momentum of the virtual photon be $k^\nu$. Similar to the real photon case in eq. (\ref{4ek}) the propagator $D_{\mu \nu}(k)$ of the virtual photon can be written as $D_{\nu \lambda}(k)=\frac{g_{\nu \lambda}}{k^2} = D_{\nu \lambda}^{physical}(k)+D_{\nu \lambda}^{pure-gauge}(k)$ where \cite{gy}
\bea
D_{\nu \lambda}^{physical}(k)=\frac{1}{k^2}[g_{\nu \lambda}-\frac{k_1\cdot k_2}{(k_1 \cdot k)(k_2 \cdot k)}k_\nu k_\lambda],~~~~~~~~~~ D_{\nu \lambda}^{pure-gauge}(k)=\frac{k_1\cdot k_2}{(k_1 \cdot k)(k_2 \cdot k)}\frac{k_\nu k_\lambda}{k^2}. \nonumber \\
\label{6ek}
\eea
The eikonal Feynman rule for the infrared divergence due to the virtual photon exchange between two electrons is given by (see eq. (4.12) of \cite{nkjhp})
\bea
\frac{k_1^\nu}{k_1 \cdot k} D_{\nu \lambda}(k) \frac{k_2^\lambda}{k_2 \cdot k}.
\label{7ek}
\eea
Using eq. (\ref{6ek}) in (\ref{7ek}) we find in the infrared limit (in the limit $k \rightarrow 0$)
\bea
\frac{k_1^\nu}{k_1 \cdot k} D_{\nu \lambda}^{physical}(k) \frac{k_2^\lambda}{k_2 \cdot k}=0,~~~~~~~~~~\frac{k_1^\nu}{k_1 \cdot k} D_{\nu \lambda}^{pure-gauge}(k) \frac{k_2^\lambda}{k_2 \cdot k}\rightarrow \infty.
\label{8ek}
\eea
Similar to the real photon case in the previous subsection we find from eq. (\ref{8ek}) that for the photon interacting with more than one electron the infrared divergence in quantum field theory can be studied by using the eikonal approximation by using the pure gauge field.

Since the light-like eikonal line produces pure gauge field in quantum field theory, see subsection \ref{eikpg}, we find that the factorization of the infrared divergence in quantum field theory due to the presence of the light-like eikonal line is enormously simplified by using the pure gauge field.

\subsection{ Light-Like Eikonal Line and The Pure Gauge Field }\label{eikpg}

In the classical mechanics the light like charge produces pure gauge field at every space-time point except at the position perpendicular to the direction of motion of the charge at the time of closest approach \cite{cs,nkjh,nke}. This property is also true in quantum field theory which can be seen as follows.

By using the path integral formulation of the quantum field theory we find that the effective lagrangian density ${\cal L}(x)$ of the photon field in the presence of light-like eikonal line is given by \cite{nayaknr}
\bea
{\cal L}(x) = 0,~~~~~~~~~~~~~~~~~~~~~~l\cdot x=0.
\label{9ek}
\eea
Similarly by using the path integral formulation of the quantum field theory we find that the effective interaction lagrangian density ${\cal L}_{int}(x)$ of the photon field in the presence of (light-like or non light-like) non-eikonal line of four velocity $v^\nu$ and the light-like eikonal line is given by \cite{nayaknr}
\bea
{\cal L}(x) = 0,~~~~~~~~~~~~~~~~~~~~~v\cdot x=0,~~~~~~~~~~~~l\cdot x=0.
\label{10ek}
\eea
Hence from eqs. (\ref{9ek}) and (\ref{10ek}) we find that the light like eikonal line produces pure gauge field at every space-time point except at the position perpendicular to the direction of motion of the eikonal line at the time of closest approach which agrees with the corresponding result in the classical mechanics \cite{cs,nkjh,nke}.

\section{Proof of color octet NRQCD factorization of P-wave heavy quarkonium production at all orders in coupling constant}

In this section we will prove the color octet NRQCD factorization of P-wave heavy quarkonium production at all orders in coupling constant at high energy colliders. In NRQCD an ultra violet cut-off $\sim M$ is introduced because of which the ultra violet behavior in NRQCD is different from the ultra violet behavior in QCD. However, the infrared (IR) behavior in NRQCD and the infrared behavior in QCD remains same. Hence the proof of the factorization of infrared divergence in a specific process in NRQCD is same in QCD. To prove factorization of infrared divergence at all orders in coupling constant it is useful to consider the path integral formulation of QCD.

In the path integral formulation of QCD the non-perturbative correlation function of the type $<\Omega| {\bar \Psi}(y) {\overline { \nabla}}_y T^a \Psi(y) a^\dagger_H \cdot a_H {\bar \Psi}(z) T^a{\overline { \nabla}}_z \Psi(z) |\Omega>$ is given by \cite{mt,abt}
\bea
&&<\Omega| {\bar \Psi}(y) {\overline { \nabla}}_y T^a \Psi(y) a^\dagger_H \cdot a_H {\bar \Psi}(z) T^a{\overline { \nabla}}_z \Psi(z) |\Omega>=\int [dQ] [d{\bar \psi}_1][d\psi_1][d{\bar \psi}_2][d\psi_2] [d{\bar \psi}_3][d\psi_3][d{\bar \Psi}][d\Psi]\nonumber \\
&&\times {\bar \Psi}(y) {\overline { \nabla}}_y T^a \Psi(y) a^\dagger_H \cdot a_H {\bar \Psi}(z) T^a{\overline { \nabla}}_z \Psi(z) ~{\rm det}[\frac{\delta \partial^\lambda Q_\lambda^h}{\delta \omega^c}]~{\rm exp}[i\int d^4x [-\frac{1}{4}F_{\nu \sigma}^s[Q(x)] F^{\nu \sigma s}[Q(x)] \nonumber \\
&&-\frac{1}{2\alpha} [\partial^\sigma Q_\sigma^s(x)]^2 +{\bar \psi}_1(x)[i{\not \partial} -m_1 +gT^s{\not Q}^s(x)]\psi_1(x)+{\bar \psi}_2(x)[i{\not \partial} -m_2 +gT^s{\not Q}^s(x)]\psi_2(x)\nonumber \\
&& +{\bar \psi}_3(x)[i{\not \partial} -m_3 +gT^s{\not Q}^s(x)]\psi_3(x)+{\bar \Psi}(x)[i{\not \partial} -M +gT^s{\not Q}^s(x)]\Psi(x)]]
\label{11ek}
\eea
where $Q_\sigma^s(x)$ is the (quantum) gluon field, the operator ${\overline { \nabla}}$ is defined in eq. (\ref{fnaq}), the parameter $\alpha$ is the gauge fixing parameter, $\Psi$ is the heavy quark field with heavy quark mass $M$, the field $\psi_k$ is the light quark field with light quark mass $m_k$ where 1, 2, 3 = u, d, s are the up, down, strange quark and
\bea
F_{\nu \sigma}^s[Q(x)]=\partial_\nu Q_\sigma^s(x) - \partial_\sigma Q_\nu^s(x)+gf^{sdc} Q_\nu^d(x) Q_\sigma^c(x).
\label{12ek}
\eea
Note that there is no ghost field in eq. (\ref{11ek}) because we directly work with the ghost determinant ${\rm det}[\frac{\delta \partial^\lambda Q_\lambda^h}{\delta \omega^c}]$ in this paper.

We have seen in the last section that the infrared divergence in the quantum field theory due to the presence of light-like eikonal line can be studied by using the pure gauge field. In QCD the pure gauge background field $A_\sigma^s(x)$ to describe the infrared divergences due to the presence of light-like eikonal line is given by \cite{nayaknr}
\bea
igT^sA_\sigma^s(x) = [\partial_\sigma \Phi(x)]\Phi^{-1}(x)
\label{13ek}
\eea
where the light-like gauge link $\Phi(x)$, see eq. (\ref{enaq}), is given by \cite{lwil}
\bea
\Phi(x)=e^{igT^s\omega^s(x)}={\cal P} e^{-ig T^d \int_0^\infty dz l\cdot A^d(x+lz)},~~~~~~~~~~~~~~l^2=0.
\label{wil}
\eea

In the path integral formulation of the background field method of QCD the non-perturbative correlation function of the type $<\Omega| {\bar \Psi}(y) {\overline { \nabla}}_y T^a \Psi(y) a^\dagger_H \cdot a_H {\bar \Psi}(z) T^a{\overline { \nabla}}_z \Psi(z) |\Omega>_A$ is given by \cite{abt}
\bea
&&<\Omega| {\bar \Psi}(y) {\overline { \nabla}}_y T^a \Psi(y) a^\dagger_H \cdot a_H {\bar \Psi}(z) T^a{\overline { \nabla}}_z \Psi(z) |\Omega>_A=\int [dQ] [d{\bar \psi}_1][d\psi_1][d{\bar \psi}_2][d\psi_2] [d{\bar \psi}_3][d\psi_3][d{\bar \Psi}][d\Psi]\nonumber \\
&&\times {\bar \Psi}(y) {\overline { \nabla}}_y T^a \Psi(y) a^\dagger_H \cdot a_H {\bar \Psi}(z) T^a{\overline { \nabla}}_z \Psi(z) ~{\rm det}[\frac{\delta B^h}{\delta \omega^c}]\nonumber \\
&&\times {\rm exp}[i\int d^4x [-\frac{1}{4}F_{\nu \sigma}^s[Q(x)+A(x)] F^{\nu \sigma s}[Q(x)+A(x)] -\frac{1}{2\alpha} [B^s(x)]^2 \nonumber \\
&&+{\bar \psi}_1(x)[i{\not \partial} -m_1 +gT^s({\not Q}^s(x)+{\not A}^s(x))]\psi_1(x)+{\bar \psi}_2(x)[i{\not \partial} -m_2 +gT^s({\not Q}^s(x)+{\not A}^s(x))]\psi_2(x)\nonumber \\
&& +{\bar \psi}_3(x)[i{\not \partial} -m_3 +gT^s({\not Q}^s(x)+{\not A}^s(x))]\psi_3(x)+{\bar \Psi}(x)[i{\not \partial} -M +gT^s({\not Q}^s(x)+{\not A}^s(x))]\Psi(x)]]\nonumber \\
\label{14ek}
\eea
where the (infinitesimal) type-I gauge transformation is given by \cite{abt,tt,zt}
\bea
A'^s_\sigma(x) =A_\sigma^s(x)+gf^{sdc}A_\sigma^d(x)\omega^c(x) +\partial_\sigma \omega^s(x),~~~~~~~~~~~~~Q'^s_\sigma(x) =Q_\sigma^s(x)+gf^{sdc}Q_\sigma^d(x)\omega^c(x) \nonumber \\
\label{15ek}
\eea
and the gauge fixing term $B^s(x)$ in the background field method of QCD is given by \cite{abt}
\bea
B^s(x)=\partial^\sigma Q_\sigma^s(x)+gf^{sdc}A_\sigma^d(x)Q^{\sigma c}(x).
\label{16ek}
\eea
Similar to eq. (\ref{11ek}) there is no ghost field in eq. (\ref{14ek}) because we directly work with the ghost determinant ${\rm det}[\frac{\delta B^h}{\delta \omega^c}]$ in this paper.

Under gauge transformation the quark field transforms as
\bea
\psi'_k(x)=e^{igT^s\omega^s(x)} \psi_k(x)=\Phi(x)\psi_k(x),~~~~~\Psi'(x)=e^{igT^s\omega^s(x)} \Psi(x)=\Phi(x)\Psi(x)
\label{20ek}
\eea
where the light-like gauge link $\Phi(x)$ is given by eq. (\ref{wil}).

By changing the integration variable $Q_\sigma^s \rightarrow Q_\sigma^s -A_\sigma^s$ in eq. (\ref{14ek}) we find
\bea
&&<\Omega| {\bar \Psi}(y) {\overline { \nabla}}_y T^a \Psi(y) a^\dagger_H \cdot a_H {\bar \Psi}(z) T^a{\overline { \nabla}}_z \Psi(z) |\Omega>_A=\int [dQ] [d{\bar \psi}_1][d\psi_1][d{\bar \psi}_2][d\psi_2] [d{\bar \psi}_3][d\psi_3][d{\bar \Psi}][d\Psi]\nonumber \\
&&\times {\bar \Psi}(y) {\overline { \nabla}}_y T^a \Psi(y) a^\dagger_H \cdot a_H {\bar \Psi}(z) T^a{\overline { \nabla}}_z \Psi(z) ~{\rm det}[\frac{\delta B_f^h[Q]}{\delta \omega^c}] {\rm exp}[i\int d^4x [-\frac{1}{4}F_{\nu \sigma}^s[Q(x)] F^{\nu \sigma s}[Q(x)]\nonumber \\
&& -\frac{1}{2\alpha} [B^s_f[Q(x)]^2 +{\bar \psi}_1(x)[i{\not \partial} -m_1 +gT^s{\not Q}^s(x)]\psi_1(x)+{\bar \psi}_2(x)[i{\not \partial} -m_2 +gT^s{\not Q}^s(x)]\psi_2(x)\nonumber \\
&& +{\bar \psi}_3(x)[i{\not \partial} -m_3 +gT^s{\not Q}^s(x)]\psi_3(x)+{\bar \Psi}(x)[i{\not \partial} -M +gT^s{\not Q}^s(x)]\Psi(x)]]
\label{17ek}
\eea
where
\bea
B^s_f[Q(x)]=\partial^\sigma Q_\sigma^s(x)+gf^{sdc}A_\sigma^d(x)Q^{\sigma c}(x)-\partial^\sigma A_\sigma^s(x)
\label{18ek}
\eea
and the $Q'^s_\sigma(x)$ in eq. (\ref{15ek}) becomes
\bea
Q'^s_\sigma(x) =Q_\sigma^s(x)+gf^{sdc}Q_\sigma^d(x)\omega^c(x) +\partial_\sigma \omega^s(x).
\label{19ek}
\eea

When the background field $A_\sigma^s(x)$ is the pure gauge background field as given by eq. (\ref{13ek}) we find from eqs. (\ref{20ek}) and (\ref{19ek}) that \cite{nayaknr}
\bea
&&F_{\nu \sigma}^s[Q'(x)]F^{\nu \sigma s}[Q'(x)]=F_{\nu \sigma}^s[Q(x)]F^{\nu \sigma s}[Q(x)],~~~[d{\bar \psi}'_k][d\psi'_k]=[d{\bar \psi}_k][d\psi_k],~~~[dQ']=[dQ],\nonumber \\
&&[B^s_f[Q'(x)]^2=[\partial^\sigma Q_\sigma^s(x)]^2,~~~~~{\rm det}[\frac{\delta B_f^h[Q']}{\delta \omega^c}]={\rm det}[\frac{\delta \partial^\lambda Q_\lambda^h}{\delta \omega^c}],\nonumber \\
&&{\bar \psi}'_k(x)[i{\not \partial} -m_k +gT^s{\not Q}'^s(x)]\psi'_k(x)={\bar \psi}_k(x)[i{\not \partial} -m_k +gT^s{\not Q}^s(x)]\psi_k(x).
\label{21ek}
\eea
Since changing the unprimed integration variables to primed integration variables does not change the value of the integration we find from eq. (\ref{17ek})
\bea
&&<\Omega| {\bar \Psi}'(y) {\overline { \nabla}}_y T^a \Psi'(y) a^\dagger_H \cdot a_H {\bar \Psi}'(z) T^a{\overline { \nabla}}_z \Psi'(z) |\Omega>_A=\int [dQ'] [d{\bar \psi}'_1][d\psi'_1][d{\bar \psi}'_2][d\psi'_2] [d{\bar \psi}'_3][d\psi'_3][d{\bar \Psi}'][d\Psi']\nonumber \\
&&\times {\bar \Psi}'(y) {\overline { \nabla}}_y T^a \Psi'(y) a^\dagger_H \cdot a_H {\bar \Psi}'(z) T^a{\overline { \nabla}}_z \Psi'(z) ~{\rm det}[\frac{\delta B_f^h[Q']}{\delta \omega^c}] {\rm exp}[i\int d^4x [-\frac{1}{4}F_{\nu \sigma}^s[Q'(x)] F^{\nu \sigma s}[Q'(x)]\nonumber \\
&& -\frac{1}{2\alpha} [B^s_f[Q'(x)]^2 +{\bar \psi}'_1(x)[i{\not \partial} -m_1 +gT^s{\not Q}'^s(x)]\psi'_1(x)+{\bar \psi}'_2(x)[i{\not \partial} -m_2 +gT^s{\not Q}'^s(x)]\psi'_2(x)\nonumber \\
&& +{\bar \psi}'_3(x)[i{\not \partial} -m_3 +gT^s{\not Q}'^s(x)]\psi'_3(x)+{\bar \Psi}'(x)[i{\not \partial} -M +gT^s{\not Q}'^s(x)]\Psi'(x)]].
\label{22ek}
\eea
From eqs. (\ref{20ek}), (\ref{21ek}) and (\ref{22ek}) we find
\bea
&&<\Omega| {\bar \Psi}(y)\Phi(y) {\overline { \nabla}}_y T^a \Phi^\dagger(y) \Psi(y) a^\dagger_H \cdot a_H {\bar \Psi}(z) \Phi(z) T^a{\overline { \nabla}}_z \Phi^\dagger(z) \Psi(z) |\Omega>_A\nonumber \\
&&=\int [dQ] [d{\bar \psi}_1][d\psi_1][d{\bar \psi}_2][d\psi_2] [d{\bar \psi}_3][d\psi_3][d{\bar \Psi}][d\Psi]\nonumber \\
&&\times {\bar \Psi}(y) {\overline { \nabla}}_y T^a \Psi(y) a^\dagger_H \cdot a_H {\bar \Psi}(z) T^a{\overline { \nabla}}_z \Psi(z) ~{\rm det}[\frac{\delta \partial^\lambda Q_\lambda^h}{\delta \omega^c}]~{\rm exp}[i\int d^4x [-\frac{1}{4}F_{\nu \sigma}^s[Q(x)] F^{\nu \sigma s}[Q(x)] \nonumber \\
&&-\frac{1}{2\alpha} [\partial^\sigma Q_\sigma^s(x)]^2 +{\bar \psi}_1(x)[i{\not \partial} -m_1 +gT^s{\not Q}^s(x)]\psi_1(x)+{\bar \psi}_2(x)[i{\not \partial} -m_2 +gT^s{\not Q}^s(x)]\psi_2(x)\nonumber \\
&& +{\bar \psi}_3(x)[i{\not \partial} -m_3 +gT^s{\not Q}^s(x)]\psi_3(x)+{\bar \Psi}(x)[i{\not \partial} -M +gT^s{\not Q}^s(x)]\Psi(x)]].
\label{23ek}
\eea
From eqs. (\ref{11ek}) and (\ref{23ek}) we find
\bea
&&<\Omega| {\bar \Psi}(y)\Phi(y) {\overline { \nabla}}_y T^a \Phi^\dagger(y) \Psi(y) a^\dagger_H \cdot a_H {\bar \Psi}(z) \Phi(z) T^a{\overline { \nabla}}_z \Phi^\dagger(z) \Psi(z) |\Omega>_A\nonumber \\
&&=<\Omega| {\bar \Psi}(y) {\overline { \nabla}}_y T^a \Phi^\dagger(y) a^\dagger_H \cdot a_H {\bar \Psi}(z) T^a{\overline { \nabla}}_z \Psi(z) |\Omega>
\label{24ek}
\eea
which proves color octet NRQCD factorization of P-wave heavy quarkonium production at all orders in coupling constant at high energy colliders.

\section{ Definition of the color octet NRQCD non-perturbative matrix element of P-wave heavy quarkonium production}

From eq. (\ref{24ek}) we find that the gauge invariant definition of the color octet NRQCD non-perturbative matrix element of P-wave heavy quarkonium production consistent with the factorization of infrared divergence at all orders in coupling constant is given by
\bea
{\cal O}_H = <\Omega| \eta^\dagger(0) \Phi(0){\overline { \nabla}} T^a \Phi^\dagger(0) \zeta(0) a^\dagger_H \cdot a_H \zeta^\dagger(0) \Phi(0) T^a{\overline { \nabla}}  \Phi^\dagger(0) \eta(0) |\Omega>
\label{25ek}
\eea
which reproduces eq. (\ref{dnaq}).

Note that since the right hand side of eq. (\ref{24ek}) is independent of the light-like vector $l^\mu$ we find that the the gauge invariant definition of the color octet NRQCD non-perturbative matrix element $<\Omega| \eta^\dagger(0) \Phi(0){\overline { \nabla}} T^a \Phi^\dagger(0) \zeta(0) a^\dagger_H \cdot a_H \zeta^\dagger(0) \Phi(0) T^a{\overline { \nabla}}  \Phi^\dagger(0) \eta(0) |\Omega>$ of P-wave heavy quarkonium production consistent with the factorization of infrared divergence at all orders in coupling constant in eq. (\ref{25ek}) is independent of the light-like vector $l^\mu$ used to define the light-like gauge link $\Phi(x)$ in eq. (\ref{enaq}).

It is useful to mention here that the heavy quarkonium is an useful probe to detect quark-gluon plasma. Hence the understanding of the heavy quarkonium production mechanism at RHIC and LHC heavy-ion colliders is necessary to study the quark-gluon plasma \cite{nj1,nj2,nj3,nj4,nj5}.

\section{Conclusions}
Recently we have proved color octet NRQCD factorization of S-wave heavy quarkonium production at all orders in coupling constant at high energy colliders in \cite{nayaknr}. In this paper we have extended this to prove color octet NRQCD factorization of P-wave heavy quarkonium production at all orders in coupling constant at high energy colliders. We have found that while the color octet NRQCD S-wave non-perturbative matrix element contains two gauge-links in the adjoint representation of SU(3), the color octet NRQCD P-wave  non-perturbative matrix element contains four gauge-links in the fundamental representation of SU(3).

\end{document}